\documentclass[pre,aps,amsmath,amsfonts,amssymb,showpacs,twocolumn]{revtex4}
\usepackage{epsfig}
\usepackage{bm}
\begin{document}
\title{Absence of the discontinuous transition in the 
one-dimensional triplet creation model}
\author{Su-Chan Park}
\affiliation{Institut f\"ur Theoretische Physik, Universit\"at zu K\"oln,
Z\"ulpicher Str. 77, 50937 K\"oln, Germany}
\date{\today}
\begin{abstract}
Although Hinrichsen in his unpublished work theoretically rebutted 
the possibility of the discontinuous transition 
in one-dimensional nonequilibrium systems unless there are 
additional conservation laws, long-range interactions, 
macroscopic currents, or special boundary conditions, 
we have recently observed the resurrection of the claim that
the triplet creation model (TC) introduced by Dickman and 
Tom\'e [Phys.  Rev. E {\bf 44}, 4833 (1991)] would show the discontinuous 
transition.  By extensive simulations, however, we find that the 
one-dimensional TC does belong to the directed percolation
universality class even for larger diffusion constant than the suggested
tricritical point in the literature. 
Furthermore, we find that the phase boundary is well described
by the crossover from the mean field to the directed percolation, 
which supports the claim
that the one-dimensional TC does not exhibit a discontinuous
transition.
\end{abstract}
\pacs{05.70.Ln,64.60.De,05.70.Fh}
\maketitle
\section{\label{Sec:intro}Introduction}
Although the field theory for the tricritical phenomena
in reaction-diffusion systems was developed more than
two decades ago~\cite{OK1987A,OK1987B} (see also Ref.~\cite{J2005}), not many 
numerical studies have followed~\cite{L2006,G2006}.
One apparent reason is the numerical difficulty, but it can be soon
overcome by the increasing computing power. More seriously,
it was strongly argued that no discontinuous transition is possible in one
dimension once there are no
additional conservation laws, long-range interactions, 
macroscopic currents, or special boundary conditions~\cite{H2000C},
which rebutted the observed discontinuous
transition in one-dimensional triplet creation model (TC) 
by Dickman and Tom\'e~\cite{DT1991}. In view of the fact that large portion 
of the studies on the absorbing phase transitions
(for a review, see, e.g. Refs.~\cite{H2000,O2004,L2004}) is 
focused on the systems in one dimension, this theory presumably has kept 
researchers from being into the tricritical phenomena.

Recently, however, numerical studies in favor of the original claim
by Dickman and Tom\'e have been reported~\cite{FO2004,CF2006,MD2007}. 
If this claim rather than the theory in Ref.~\cite{H2000C} turns 
out to be right, we would observe an avalanche of studies on 
the tricritical phenomena. Unfortunately, however, no theoretical argument
regarding the mechanism to stabilize a domain in one dimension
has been suggested as yet. Moreover,
the tricritical point of the diffusion rate
reported in Ref.~\cite{FO2004,CF2006} is too large to reject the
opinion that the system will eventually crossover to the directed percolation (DP)
universality class after a long transient time.
Recent study by Ferreira and Fontanari~\cite{FF2009}
using $n$-site approximation alluded to the crossover
rather than the tricritical behavior, though they 
did not strongly put forward such a scenario 
because of the computational limitation of their method.
Actually, Hinrichsen~\cite{H2000C} numerically showed that the simulation
time in Ref.~\cite{DT1991} was too short to see the correct scaling behavior.
Interestingly, Cardozo and Fontanari~\cite{CF2006} also 
refuted the value of the tricritical point originally suggested.
Hence, if the argument by Hinrichsen~\cite{H2000C} 
is right, it is very probable that 
more extensive simulations than those in Ref.~\cite{CF2006} would
revive the history. 

Indeed, we found the DP scaling over the parameter range where 
the tricritical point was located by Cardozo and 
Fontanari~\cite{CF2006}. 
This paper is for providing numerical evidences to support the theory
suggested by Hinrichsen~\cite{H2000C}.

The rest of this paper is organized as follows:
Section~\ref{Sec:model} introduces
the $d$-dimensional TC and explains
the algorithm implemented for numerical simulations.
The numerical results showing the DP scaling behavior for larger
diffusion rate than the previously reported tricritical point will be
presented in Sec.~\ref{Sec:DP}. 
In Sec.~\ref{Sec:2D}, we will argue that there is a crossover rather than
a tricriticality in one dimension by studying 
the behavior of the phase boundary.
Section~\ref{Sec:sum} summarizes the work.

\section{\label{Sec:model}triplet creation model}
The TC is an interacting hard core particles system on a $d$-dimensional
hypercubic lattice with three processes, hopping (with rate $D$), 
spontaneous annihilation (with rate $\gamma$), and
creation by a triplet (with rates $s$)~\cite{DT1991}.
By hard core is meant that no two particles can occupy the same site 
at the same time.
By a suitable time rescaling, we can set $D+\gamma+s=1$ without
loss of generality. It is also convenient to introduce the
annihilation probability $p$ such that 
$\gamma = (1-D) p$ and $s = (1-D)(1-p)$. 
The detailed dynamics is to be explained in terms of the algorithm used
for simulations.

At time $t$, $N_t$ particles are distributed on a $d$-dimensional hypercube 
of size $L^d$ ($N_t$ is a random variable).
Each site is represented by a lattice vector ${\bm m}=(m_1,\ldots,m_d)$ 
($0\le m_i \le L-1$). The unit vector along direction $i$ is denoted by
$\bm{e}_i$ ($i=1,\ldots,d$).
In all simulations in this paper, a fully occupied initial condition and 
periodic boundary conditions are assumed. 

The algorithm begins with a random selection of a particle in the system.
For convenience, let us refer to the lattice vector of 
the site the selected particle resides as $\bm{m}$. 
With probability $D$, hopping is attempted to a target site
which is chosen randomly among $2d$ nearest neighbors of the site $\bm{m}$.
This hopping is successful only if the target site is empty (hard core
exclusion), otherwise there is no configuration change.
With probability $1-D$, either annihilation (with probability $p$)
or creation (with probability $1-p$) will be attempted.
When annihilation is decided, the selected particle will be irreversibly
removed from the system. If the creation is to occur,
one of the directions $i$ ($i=1,\ldots,d$) is selected randomly.
Two sites 
$\bm{m}+\bm{e}_i$ and $\bm{m}+2\bm{e}_i$ are checked whether
both sites are occupied or not.
If both sites are also occupied, 
one of the two sites,  $\bm{m}+3 \bm{e}_i$ or $\bm{m}-\bm{e}_i$,
is chosen at random as a target site 
and a new particle is created at the target site provided
it is empty. If any of the conditions for
creation is not satisfied, nothing happens.
After an attempt to change a configuration,
time increases by $1/N_t$ regardless of its success. The above
procedure iterates until either the system reaches the absorbing state where
no particle remains
or time gets larger than the preassigned observation time.

When $d=1$, the above dynamics is exactly the same as that in
Ref.~\cite{CF2006} with $p=1/(1+\lambda)$. This is because
the probability to find three occupied sites in a row does not depend on 
whether two sites $\bm{m}+\bm{e}_i$ and $\bm{m}+2\bm{e}_i$ (in this work)
or $\bm{m}+\bm{e}_i$ and $\bm{m}-\bm{e}_i$ (in Ref.~\cite{CF2006}) are examined.

Although we introduced $d$-dimensional TC, discussions from Sec.~\ref{Sec:DP} 
on will be focused only on the one-dimensional model. Thus, the dimensionality
of the model will not be mentioned explicitly in the following.

In the simulation, we measure the particle density $\rho(t)=\langle N_t
\rangle /L$, where $\langle \ldots \rangle$ means the average over
independent realizations,
and the (survival) probability that there is a particle in the system
at time $t$. The measurement of the survival probability is mainly for making
sure that the system is large enough not to be affected 
by the finite size effect up to the observation time.

\begin{figure}[t]
\includegraphics[width=0.44\textwidth]{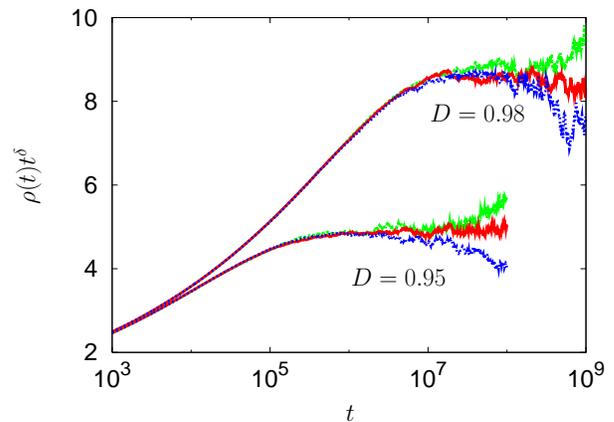}
\caption{\label{Fig:TCP} (color online) Plots of $\rho(t) t^\delta$ vs $t$ with 
$\delta = 0.1595$ (critical exponent of the DP class) for $D=0.95$ (lower 
three curves) and $D=0.98$ (upper three curves) in semi-logarithmic scales. 
The values of $p$ for
$D=0.95$ are 0.089~89, 0.089~895, and 0.0899 from top to bottom.
The values of $p$ for
$D=0.98$ are 0.094~224, 0.094~226, and 0.094~228 from top to bottom.}
\end{figure}
\section{\label{Sec:DP} Critical density decay}
This section studies the critical behavior of the TC 
with $D=0.95$ and $D=0.98$. Since the question in this section
is whether the TC shows the discontinuous transition
or the continuous transition governed by the DP fixed point, 
observing the DP scaling behavior of a single quantity is enough
for our purpose.  Anticipating the conclusion, we only
study how the density decays near criticality.

Figure~\ref{Fig:TCP} depicts 
$\rho(t) t^\delta$ as a function of $t$ in semi-logarithmic scales,
where $\delta=0.1595$ is the critical density decay exponent of the DP class
taken from Ref.~\cite{J1996}.
The system size in the simulations for $D=0.95$ ($0.98$) is 
$L=2^{18}$ ($2^{17}$). The number of independent runs
for each data set varies from $16$ ($D=0.95$ and $p=0.089~89$) to 
$100$ ($D=0.98$ and $p=0.094~226$). The system evolves up to $t=10^9$ at the 
longest and no sample falls into the absorbing state during simulations.
For $D=0.95$, the curve corresponding to $p=0.089~89$ ($0.0899$) 
veers up (down), which indicates that the system is in the active (absorbing)
phase. At $p=0.089~895$, the curve is flat for more than two log-decades.
Hence we conclude that the TC with $D=0.95$ belongs to the DP class
with critical point $p_c = 0.089~895(5)$, where the number in parentheses
indicates the error of the last digit. If we write the critical point
using $\gamma = (1-D)p$, we get $\gamma_c=0.004~4948(3)$ which should
be compared with 0.004~50(1) reported in Ref.~\cite{CF2006}.
One should note that the DP scaling is observable from $t=10^6$ which 
is the end point of the simulation for $D=0.95$ in Ref.~\cite{CF2006}
(see Fig.~3(b) of Ref.~\cite{CF2006}). Likewise, the simulation
results for $D=0.98$ show the similar behavior as those for $D=0.95$ (see 
upper three curves in Fig.~\ref{Fig:TCP}). 
The critical point for $D=0.98$ is found to be
$p_c = 0.094~226(2)$ or $\gamma_c = 0.001~888~452(4)$.  In Ref.~\cite{CF2006},
the critical value $\gamma_c$ for $D=0.98$ was reported as 
0.001~886(2) and the simulation was terminated around $t=10^7$ from when
the DP scaling is observable. 

To conclude this section, the TC up to $D = 0.98$ belongs to the DP class
and previous claim of the existence of the tricritical point below $D=0.98$
is refuted. Our results also explain why Cardozo and 
Fontanari~\cite{CF2006} observed continuously varying exponents 
as well as the compact growth; the system was analyzed 
before the correct scaling behavior was detected.

\section{\label{Sec:2D} Crossover from the mean field to the directed percolation}
In Sec.~\ref{Sec:DP}, we numerically confirmed that up to $D = 0.98$ 
the TC does show continuous transition governed by the DP fixed point. 
Although this refuted the previous claim~\cite{FO2004,CF2006}
that the transition nature changes at a certain $D$ smaller than $0.98$, 
the possibility of the discontinuous transition at $0.98 <D <1$ is
still open. To provide an evidence that the DP fixed point governs the
critical behavior beyond $D=0.98$, this section studies the phase 
boundary near $D=1$ with the focus on a possible crossover. 

The investigation of the phase boundary to settle a controversy is not 
without precedent.  The present author and his collaborator 
tried to resolve the controversy around the pair contact process
with diffusion (for a review, see, e.g., Refs.~\cite{HH2004,PP2008EPJB}) 
by studying the phase boundary of the crossover models~\cite{PP2006}.
Although this study could not elicit a full consensus, 
it certainly gives a hint about the system. 
So we think it is worth while to investigate the phase boundary
of the TC.  But, in Sec.~\ref{Sec:DP}, 
we have presented the simulation results only for
$D=0.95$ and $D=0.98$, which is certainly not enough to
see the structure of the phase boundary. For a better bird's-eye view,
we will include some other critical points for $D<0.95$ even though
we will not be presenting details.

Before directly analyzing the  phase boundary, 
we will discuss some features of it which can be inferred
without resorting to the nature of the transition.
First, the transition point $p_c(D)$ is argued to
be an increasing function of $D$. Second,
$p_c(D)$ is expected to approach to the transition point of the mean field 
(MF) theory as $D\rightarrow 1$.
By the MF theory in this paper is 
exclusively meant the one-site approximation
(for a detail, one may consult Sec. 3 of Ref.~\cite{FF2009}) 
\begin{equation}
\frac{d \rho}{d\tau} = -p \rho + (1-p)\rho^3(1-\rho),
\label{Eq:MF}
\end{equation}
which exhibits a discontinuous transition with the transition
point $p_0 = \frac{4}{31}$.

To argue that $p_c(D)$ increases with $D$,
it is convenient to modify the TC by
rescaling time $\tau \equiv (1-D) t$. 
Note that for $D<1$, however close $D$ is to 1, 
the steady state property of the modified model is identical to 
the original one. In this modified model,
a single particle dies out with rate $p$ and a triplet attempts to
branch an offspring with rate $1-p$. Obviously, these rates are not dependent
on $D$. The diffusion rate is now
$\tilde D \equiv D/(1-D)$ which is an increasing function of $D$. 
Since the annihilation occurs regardless of the environment
of a to-be-annihilated particle, the diffusion 
cannot directly affect the particle number fluctuation due to the annihilation.
On the other hand, the branching is influenced by the diffusion,
since a triplet can be either broken or newly formed by the movement
of particles. 
Hence, whether the diffusion enhances or reduces the activity
of the system for given $p$ can be answered by understanding if 
the diffusion will increase or decrease the number of
active triplets, namely, triplets with a vacant neighbor.

Due to the hard core exclusion, a cluster of particles will lose
one by the diffusion only at {\em boundaries}. However, even if a 
cluster loses a single particle at a boundary, the number of active triplets 
does not decrease if the size of the cluster is larger than 3 (it can
even increase the number of active triplets if the configuration
change due to the diffusion is like $01101111 \rightarrow 01110111$, where
1 stands for a particle and 0 for a vacancy). 
Thus, the reduction of the number of active triplets 
by the diffusion occurs in very restricted situations.
On the other hand, the diffusion can mediate the formation of (active)
triplets in the {\it bulk} of region where none exists.
To sum up, the diffusion tends to enhance the activity, which entails
that the transition point should increase with $D$.

Next, we will discuss the limiting value of $p_c(D)$ as $D\rightarrow 1$,
which will be denoted by $\tilde p$.  To find $\tilde p$, we start from 
arguing that the TC under the $D\rightarrow 1$ limit is deeply related to the 
MF theory. This connection makes sense only when time is suitably rescaled as 
before. With this time scale, the diffusion rate $\tilde D$ grows indefinitely
as $D\rightarrow 1$, but the annihilation and creation rates remain finite. 
From now on, we will call the limit $D \rightarrow 1$ the fast diffusion limit
and the modified TC with rescaled time is always assumed when we
are discussing the fast diffusion limit.

A finite system under the fast diffusion limit can be interpreted 
as follows: Right after any reaction (either annihilation or creation), 
the system arrives at the steady state of 
the diffusion process in no time and remains there until another 
reaction occurs. 
In the steady state of the diffusion-only system, 
all possible configurations for given number of particles have equal 
probability. Hence, the probability that a site is vacant and its
three consecutive neighbors are occupied at time $\tau$
is $(L-n)(n)_3/(L)_4$, where $L$ is the system size and $n$ is the number of particles in the
system at time $\tau$, and $(m)_k \equiv m(m-1)\ldots(m-k+1)$.
Of course, the probability that a site is occupied at time $\tau$ 
is $n/L$. Since these probabilities do not depend on where the
site is located, the probability distribution is fully
specified by the number of particles and the master equation of
the TC under the fast diffusion limit is reduced to
\begin{equation}
\partial_\tau P_n = 
a_{n+1} P_{n+1} + c_{n-1} P_{n-1} - (a_n + c_n) P_n,
\label{Eq:MF_finite}
\end{equation}
where $a_n \equiv p n$, $c_n \equiv (1-p)(L-n)(n)_3/(L-1)_3$, and
$P_n$, though the argument of it is not written explicitly, 
is the probability that there are $n$ particles at time $\tau$.
For convenience, we set $P_{L+1}= P_{-1} = 0$.

The meaning of $a_n$ and $c_n$ in Eq.~\eqref{Eq:MF_finite} can be
interpreted as follows: 
$a_n$ means each particle (there are $n$ particles) dies with rate $p$
and $c_n$ means that each vacant site (there are $L-n$ vacant sites)
becomes occupied with rate $(1-p)(n)_3/(L-1)_3$, which is the
dynamic rule of the TC on a fully connected graph.
Hence, the master equation of the TC on a fully connected graph
is exactly Eq.~\eqref{Eq:MF_finite}. If we take the thermodynamic limit
($L\rightarrow \infty$) to Eq.~\eqref{Eq:MF_finite},
the law of large numbers has the density follow Eq.~\eqref{Eq:MF}, that is, 
the MF theory. Now it is clear why we used $\tau$ in
Eq.~\eqref{Eq:MF} as a time parameter.
For convenience and because of an obvious reason,
we will refer to the TC on a fully connected graph with finite size as
the MF model.

\begin{figure}[t]
\includegraphics[width=0.44\textwidth]{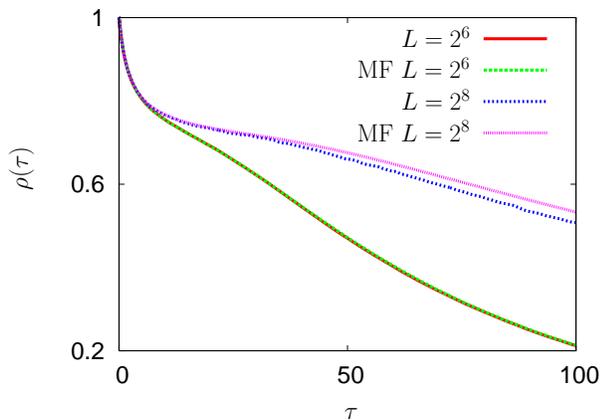}
\caption{\label{Fig:MF_small} (color online) Comparison of the numerical solutions of
Eq.~\eqref{Eq:MF_finite} with the simulation results of the 
TC with $1-D=10^{-4}$. As explained in the text, the time is rescaled as $\tau 
= (1-D) t$. The systems sizes are $L=2^6$ (two below curves, though indiscernible)  and $2^8$ (two above curves). $p$ is fixed at $\frac{1}{8}$.
For $L=2^6$, no difference is detectable between simulation and the MF
solution. On the other hand, the system with $L=2^8$ 
is distinct from the MF solution after $\tau = 50$ 
(MF solution is slightly above the simulation results).
}
\end{figure}

Although the fast diffusion limit was taken to arrive at the MF model
in the above discussion, 
one can observe the behavior of the MF model even for nonzero $1-D$
once $(1-D)L^2 \ll 1$~\cite{DT1991}.  In Fig.~\ref{Fig:MF_small},
we compare the simulations of the TC for $1-D=10^{-4}$ 
to the (numerical) solutions of the MF model at
$p=\frac{1}{8}$. The system sizes are $L=2^6$ and $L=2^8$. 
For the solutions of the MF model, we numerically integrate
Eq.~\eqref{Eq:MF_finite} with the initial 
condition $P_n(\tau = 0) = \delta_{nL}$ ($\delta$ here
is the Kronecker delta symbol).
The simulation results are not discernible from the behavior
of the MF model for $L=2^6$ [$(1-D)L^2 \approx 0.4$], but 
clear deviation is observed when $L=2^8$. 

In the above discussion, the thermodynamic limit, when necessary, 
is preceded by the fast diffusion limit.
However, what we are interested in is the behavior of the 
TC with infinite size under the fast diffusion limit. That is,
the thermodynamic limit should be taken before the fast diffusion limit.  
When two limits are involved in the calculation, one should be careful
about which limit is taken first.  Fortunately, 
the order of two limits are irrelevant in most cases.
What is meant by most cases will become clear in due course.

\begin{figure}[t]
\includegraphics[width=0.44\textwidth]{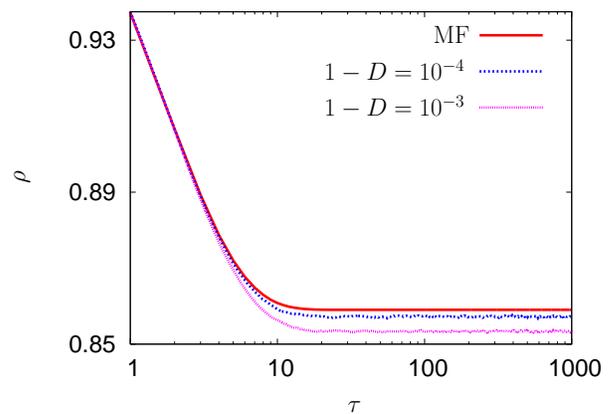}
\caption{\label{Fig:MF_finite} (color online) Comparison of the numerical solutions of
Eq.~\eqref{Eq:MF} with the simulation results of the TC with $1-D=10^{-3}$ and $1-D = 10^{-4}$ at $p=0.094~226$. 
As in Fig.~\ref{Fig:MF_small}, $\tau$ is the rescaled time. 
The systems size for simulations is $L=2^{17}$. 
As $1-D$ decreases, the density approaches to the MF solution.
}
\end{figure}
Our discussion commences with the limiting behavior of the TC 
in the active phase ($p < \tilde p$).
Let the critical diffusion probability for given $p$ be 
denoted by $D_c(p)$. Since $p_c(D)$ is a monotonous function,
$D_c(p)$, if exists, is uniquely determined.
If $1> D > D_0$, where $D_0$ is an arbitrary number strictly
larger than $D_c(p)$, the system is in the active phase
and the correlation length is bounded for all $D$ in this regime.
If the system size is much larger than the bound 
of the correlation lengths in this regime, the behavior in the thermodynamic
limit is observable for any $D_0<D<1$ even though the system is finite.
Since we can always choose $D$ such that
$(1-D)L^2 \ll 1$ for given $L$, the TC under the fast diffusion limit
preceded by the thermodynamic limit is identical to the MF theory
at the same value of $p$.

To support this argument, Fig.~\ref{Fig:MF_finite} 
compares the MF theory with simulations
of the TC for $D=0.999$ and $D=0.9999$
at $p = 0.094~226$ which is the critical point for $D=0.98$.
Obviously, the density of the TC obtained from simulations approaches 
to the MF theory as $D \rightarrow 1$.

The above consideration reveals that $\tilde p$ should not
be larger than $p_0$, otherwise the steady state density
for $p_0 < p <\tilde p$ would decrease with $D$, which is contradictory
to the role of the diffusion as a enhancer of the branching.

If $p>p_0$ where the TC as well as the MF model is in the absorbing phase, 
we can arrive at the same conclusion
as above, that is, the behavior of the TC under the fast diffusion limit 
is not affected by whether before or after the thermodynamic limit is taken.

To complete our discussion, we first have to figure out if 
$\tilde p$ can be strictly smaller 
than $p_0$.  Actually, this possibility was suggested by Fiore and de 
Oliveira~\cite{FO2004} by extrapolating the phase boundary which
was obtained numerically. 
Note that the discontinuity of a phase boundary {\it per se} is 
not an unrealistic conclusion.  One can even find another report which, 
though in a different context, shows the discontinuity of the phase 
boundary~\cite{PP2007}. 

\begin{figure}[t]
\includegraphics[width=0.45\textwidth]{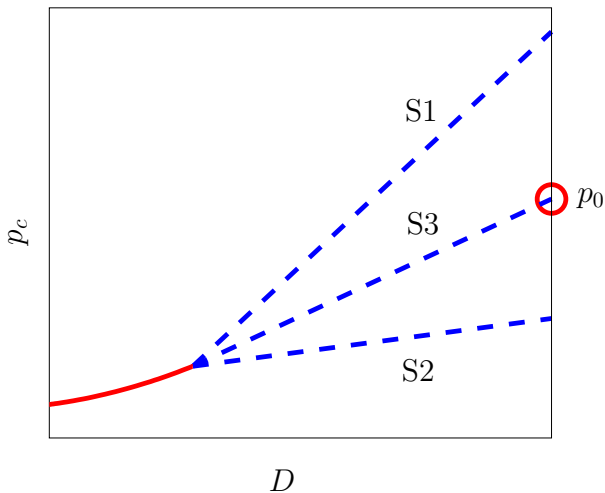}
\caption{\label{Fig:scenarios} (color online) 
Schematic representation of three scenarios
regarding the limiting value of $p_c$ as $D\rightarrow 1$. 
$p_0$ indicated by the open circle 
is the mean field transition point and scales
in this figure are arbitrary.
The broken lines depict the anticipated phase boundaries of each scenarios
with known transition points depicted by the solid curve.
The limiting value $\tilde p$ of each scenario is the point where 
the broken line meets the vertical line with $p_0$ on it. 
The first scenario (S1) is not acceptable 
because the steady state density should
increase as $D$ increases (see the text). The second scenario (S2) which
was proposed in Ref.~\cite{FO2004} is possible only when
there is a line of critical points (see the text), 
which does not seem plausible. 
Hence the last scenario (S3) is concluded to be the right one.
}
\end{figure}

However, this scenario does not seem plausible.
To demonstrate why it is not likely, let us think about the situation where
$\tilde p < p <p_0$, $(1-D) L^2 \ll 1$, and $L$ is much larger than
the correlation length for given $p$ and $D$. 
As shown before, the system with $(1-D)L^2 \ll 1$ is well
described by the MF model. Since $L$ is assumed very large and this
value of $p$ corresponds to the active phase of the MF theory,
the MF model, and accordingly the TC itself with above mentioned parameters, 
stays at the steady state of the MF theory with nonzero density 
for long time of the order of $O(\exp(L))$ (note that time in this context
is the rescaled time $\tau$). 
On the other hand, the system size is assumed much larger than the correlation
length and the system is in the absorbing phase by assumption, 
so the density should decay exponentially after time of the
order of $O(L^0)$, 
which is contradictory to the previous consideration.
Of course, if the correlation length in this regime diverges
faster than $(1-D)^{-1/2}$ as $D\rightarrow 1$, the assumption
$(1-D)L^2 \ll 1$ with $L$ larger than the correlation length 
is not valid. However, this scenario of the 
diverging correlation length in this regime also does not seem plausible, 
because it suggests that the correlation length should diverge for 
$\tilde p<p<p_0$ and should become finite as soon as $p>p_0$. 
Hence, we conclude that $\tilde p$ is equal to $p_0$.
The above discussion is summarized in Fig.~\ref{Fig:scenarios}.

Up to now, we have not resorted to any assumptions about the transition
nature. If we assume that the transition
is always continuous, what can we say about the limiting behavior
of the TC at $p=\tilde p=p_0$. In this case, the correlation 
length diverges as $(1-D)^{-\nu_\perp}$, where $\nu_\perp \simeq 1.09$
is the critical exponent of the DP class. Accordingly, 
it is not possible to think about the system size which is much larger
than the correlation length but $(1-D)L^2 \ll 1$ when $D$ is very 
close to 1.  That is, the fast diffusion and thermodynamic limit do not commute
at $p=p_0$. However, 
even if we assume that there is a discontinuous transition for finite $D$,
the noncommutability of two limits at $p=p_0$ is still applicable, because
the steady state density for $p=p_0$ is zero for any value of $D<1$,
though the MF theory has a finite density at $p=p_0$.
In a sense, this noncommutability is originated from the fact
that the MF theory exhibits the discontinuous transition which is
characterized by the discontinuity of the density at stationarity.
At any rate, the thermodynamic limit commutes with the fast diffusion limit 
in most cases except at $p=p_0$.

\begin{figure}[t]
\includegraphics[width=0.44\textwidth]{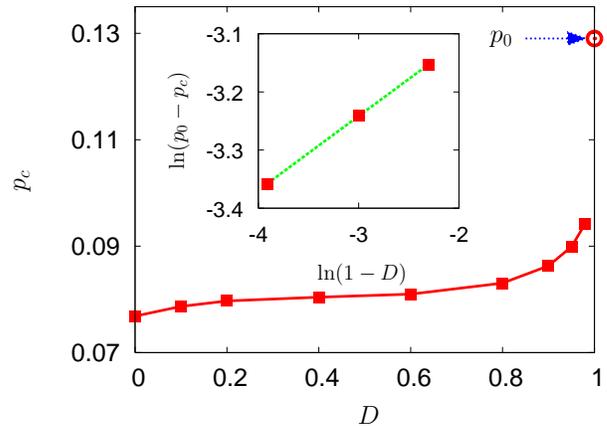}
\caption{\label{Fig:pb} (color online) Plot of $p_c$ as a function of $D$ for
the TC. The MF transition point $p_0$ is indicated
by an arrow. Inset: Plots of $\ln(p_0-p_c)$ vs $\ln(1-D)$
and its fitting function (see text).  Symbols are from the simulations and
the straight line is from the fitting.
}
\end{figure}
Although the above consideration reveals that the phase boundary should
approach to the MF transition point, it does not give any information about the
transition nature.  To get a nontrivial conclusion, the numerically
obtained phase boundary will be examined.

Figure~\ref{Fig:pb} depicts the phase boundary in 
$D-p$ plane. As argued before, the transition point increases
with $D$. In the range $0 \le D \le 0.8$ where no controversy has ever been 
raised, the critical points do not change much, compared to the change from
$D=0.8$ to $D=0.98$. It is likely that the phase boundary 
approaches to the  MF transition point with infinite slope.
If we fit the phase boundary using the fitting function
\begin{equation}
\ln(p_0 - p_c) =  \frac{1}{\phi}\ln(1-D) + b,
\label{Eq:crossover}
\end{equation} 
with two fitting parameters $\phi$ and $b$, 
we get $p_c \approx p_0 - 0.057(1-D)^{0.127}$ 
($\phi \approx 8$) from
last three points [without $p_0$ and with $p_c= 0.08630(1)$ for $D=0.9$;
see Inset of Fig.~\ref{Fig:pb}].
Although the accuracy of the fitting should not be exaggerated, the clean
power-law behavior shown in Inset of Fig.~\ref{Fig:pb}
strongly suggests that the phase boundary approaches to the MF transition point
with infinite slope. Note that this infinite slope is
the characteristics of the crossover behavior~\cite{LS1984}. 
Also note that  a ``crossover'' within a single universality class, which
is actually not a crossover at all, 
does not show such a singular behavior~\cite{PP2007}. 
Although we are considering the discontinuous transition,
it is a natural generalization of the claim in Ref.~\cite{PP2007}
that no singularity can appear in the phase boundary 
along which only discontinuous transitions occur.
Thus, the singular behavior of the phase boundary
near $D=1$ supports that no discontinuous transition exists
for $D<1$.

\section{\label{Sec:sum}Summary and Discussion}
In summary, we investigated the triplet-creation model (TC) 
in one dimension.  By extensive numerical 
simulations, we refuted the previous estimation
of the tricritical point~\cite{FO2004,CF2006}.
We only observed the directed percolation scaling up to $D=0.98$
which is larger than the tricritical point suggested in 
Ref.~\cite{FO2004,CF2006}.
To go further beyond $D>0.98$, we analyzed the phase boundary
near $D=1$. At first, we argued that the phase boundary should approach
to the mean field transition point as $D\rightarrow 1$.
Using this information, the phase boundary was analyzed using the power law
fitting function Eq.~\eqref{Eq:crossover} to find $\phi \simeq 8$,
which, according to the general theory of the crossover~\cite{LS1984},
strongly suggests the absence of discontinuous transitions in one dimension.

As a final remark, we would like to comment on the conserved ensemble
of the TC. In Ref.~\cite{FO2004}, Fiore and de Oliveira
studied the conserved ensemble of the TC. They convincingly argued
that the conserved version is equivalent to the TC studied in
this paper, based on the proof in Refs.~\cite{HW2002,O2003}.
This argument also embraces the limiting case ($D\rightarrow 1$);
see Sec. III in Ref.~\cite{FO2004}. Actually, the equivalence becomes
trivially true for the MF model under the thermodynamic limit
because the equivalence criterion which is Eq. (30) of  Ref.~\cite{O2003} 
becomes identical to the steady state condition of Eq.~\eqref{Eq:MF}.
Thus, the direct comparison of the results in this paper with those
in Ref.~\cite{FO2004} is fully legitimate although we studied
the different ensemble.  In this respect, this work clearly suggests a caveat.
Unlike the belief regarding the merit of the conserved ensemble (for example,
see Sec. V in Ref.~\cite{FO2004}), one should study the conserved version
with caution even if the discontinuous transition is the main interest.

\section*{Acknowledgment}
S.-C.P. would like to thank the hospitality of Korea Institute for Advanced
Study where this project was planned during his visit in 2007.
This work has been supported by DFG within SFB 680 
\textit{Molecular Basis of Evolutionary Innovations.}

\end{document}